\documentclass[aps,prb,twocolumn,groupedaddress]{revtex4}
\usepackage{upgreek}

\usepackage{graphicx}   % For eps figures
\usepackage{epsfig}     % Alternative package

\begin{document}

\title{Circuit QED with a Flux Qubit Strongly Coupled \\to a Coplanar Transmission Line Resonator}
%\submitto{SUST}

\author{T.Lindstr\"{o}m$^1$, C.H. Webster$^1$, J.E. Healey$^2$, M. S. Colclough$^2$  C.M.Muirhead$^2$, A.Ya.Tzalenchuk$^1$}
\address{$^1$National Physical Laboratory, Hampton Road, Teddington, TW11 0LW, UK}
\address{$^2$ University of Birmingham, Edgbaston, Birmingham B15 2TT, UK}

\date{\today}

\begin{abstract}

We propose a scheme for circuit quantum electrodynamics with a
superconducting flux-qubit coupled to a high-Q coplanar resonator.
Assuming realistic circuit parameters we predict that it is possible
to reach the strong coupling regime.  Routes to metrological
applications, such as single photon generation and quantum
non-demolition measurements are discussed.
\end{abstract}


%\submitto{SUST}
\maketitle

\section{Introduction}

Until a few years ago it was an open question whether true quantum
effects such as quantum entanglement would ever be observed in a
man-made macroscopic electronic device. However, over the past
decade, quantum coherence has been demonstrated in a variety of
macroscopic systems, including superconducting circuits
\cite{nakamura, vanderwal, chiorescu, martinis2003}. Many of these
experiments drew inspiration from the pioneering work on atomic
qubits that took place a decade earlier \cite{Zeilinger99}. As the
fields of atomic physics and quantum optics continue to advance,
it makes sense to continue to look to them for guidance.

The universal nature of quantum mechanics is greatly to our
advantage, in that the terminology and methodology apply as well
to macroscopic as to microscopic systems.  This allows well-known
results from atomic physics and quantum optics to be used to plan
and predict the outcome of experiments on solid state devices.
Recently, such techniques have been applied with great success to
implement a number of ideas such as quantum state tomography
\cite{steffen}, Mach-Zehnder interferometry \cite{oliver} and
sideband cooling\cite{valenzuela}.

This approach has also been very successful in the field of cavity
quantum electrodynamics (CQED) [see Ref.\ \cite{gerry} for an
introduction]. In CQED an atomic 2-level system  (i.e. a qubit) is
made to interact with a high-finesse optical cavity with a
coupling energy $\hbar g$. Provided that the relaxation rates
$\gamma$ of the qubit and $\kappa$ of the cavity field are smaller
than $g$ (known as the strong coupling criterion), it is possible
to observe a coherent exchange of energy between the qubit and the
cavity field. The resulting entangled states can be detected
spectroscopically. Recently,  Schoelkopf  and co-workers
\cite{wallraff,blais}  achieved strong coupling in a macroscopic
circuit comprising a superconducting charge qubit and a coplanar
transmission line resonator. This new field is known as
circuit-QED, and has many potential applications such as the
generation and detection of single microwave photons. More
recently strong coupling was observed in experiments on photonic
crystals \cite{yoshie} and quantum dots\cite{reithmaier}.

Much of the work to date on qubit-cavity systems has been focused
on methods for making quantum non-demolition (QND) measurements of
the qubit state. QND schemes have been used to read-out single
qubits \cite{wallraff,lupascu} and also to measure the photon
number in the cavity \cite{schuster}. Several ways of reading out
qubits using low-$Q$ cavities/resonators have also been reported
\cite{Johanson, ilichev}, but these do not allow the strong
coupling regime to be reached.

  The benefits of
superconducting circuit-QED systems over atomic systems are
twofold. Firstly, the qubit energy can be tuned by varying the
external magnetic field, enabling control over the qubit-cavity
interaction. Secondly, qubit parameters such as the level
separation at the degeneracy point can be engineered through
appropriate circuit design.  The main advantage of flux qubits
over other types of superconducting qubits is that they are less
susceptible to fluctuations in the background charge and the
associated noise. This makes them less prone to decoherence, and
therefore easier to manipulate in a deterministic way.

In section II we summarise the established background theory in
the context of our proposed system; in III we show that it is
possible to achieve strong coupling between a superconducting flux
qubit \cite{Mooij-1999} and a high-Q coplanar transmission line
resonator; in IV we present computer simulations of the resonator
response; and in V we propose a scheme for producing single
photons on demand.

\section{Theory of Circuit QED with a Flux Qubit}

\begin{figure}[t]
\centering
\includegraphics[width=7cm]{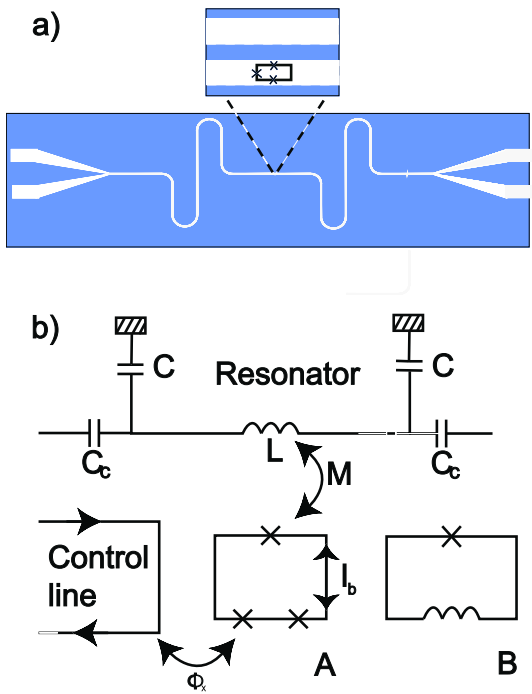}
\caption{\label{fig:CQED_setup} a) Sketch of a typical coplanar
waveguide resonator of length l=$\lambda/2\approx11$ mm. Shown is
also how the qubit can be placed in between the centre conductor
and the ground plane of the waveguide. b) Schematic diagram of a
superconducting qubit coupled to a coplanar transmission line
resonator.  (A) Persistent current qubit.  (B) RF SQUID.  $M$ is
the mutual inductance between the qubit and resonator, and
$\Phi_X$ is the magnetic flux threading the qubit loop.}
\end{figure}

In this section we analyze the processes which occur when a
superconducting qubit is coupled to a superconducting coplanar
transmission line resonator, as shown in Fig.\
\ref{fig:CQED_setup}. The effect of coupling a flux qubit to a
resonator has previously been experimentally demonstrated
\cite{chiorescunature} but the quality factor of the resonator was
too small to fulfill the strong coupling criteria.

Two types of flux qubits will be discussed - an RF SQUID, which
consists of a single Josephson junction in a superconducting loop,
and a persistent current qubit (PCQ), which has three junctions in
the loop, one of which is smaller than the others by a factor
$\alpha$ \cite{Mooij-1999}. Both need to be biased by an external
magnetic flux $\Phi_x$, which tunes the energy level separation
through an anticrossing at $\Phi_x = 0.5\Phi_0$. The resonator is
a coplanar transmission line with inductance $L$ and capacitance
$C$, which is weakly coupled to external transmission lines via
coupling capacitors $C_C$.
The Hamiltonian $H$ of the complete system is
\begin{equation}
H=H_q+H_r+H_g+H_I+H_E.
\label{eq:hamil}
\end{equation}
$H_q$ describes the qubit, $H_r$ the resonator, and $H_g$ the
interaction between them.  $H_I$ denotes the interaction of the
system with a periodic drive field. Finally, $H_E$ describes the
interaction of the resonator with its environment, resulting in the
loss of photons to the external transmission lines and the interaction of the
qubit with its environment, resulting in spontaneous decay from the
excited state to the ground state.

The qubit Hamiltonian $H_q$ is given by the expression
$(\hbar/2)(-\epsilon \sigma_z-\delta \sigma_x)$ where $\sigma_z$
and $\sigma_x$ are Pauli spin matrices, $\delta$ is the the level
repulsion, $\epsilon = (2I_p/\hbar) (\Phi_x-\Phi_0/2)$, and $I_p$
is the persistent current. In the case of an RF SQUID suitable for
operation as a qubit, $I_p$ is roughly equal to half the critical
current of the single junction, whereas, for the persistent
current qubit, $I_p$ is approximately equal to the critical
current of the smallest of the three junctions. If the qubit is
operated at or near the degeneracy point, $H_q$ can be expressed
more simply by transforming to the basis in which the ground state
$\vert\downarrow\rangle$ and excited state $\vert\uparrow\rangle$
correspond to symmetric and antisymmetric superpositions of
clockwise and anti-clockwise persistent currents. This yields
\begin{equation}
H_q=\frac{\hbar\omega_0}{2}(\vert\uparrow\rangle\langle\uparrow\vert-
\vert\downarrow\rangle\langle\downarrow\vert)=\frac{\hbar\omega_0}{2}\sigma_z,
\end{equation}
where the level separation is $\hbar\omega_0$, $\omega_0$  being the qubit Larmor frequency
$\sqrt{\epsilon^2+\delta^2}$.

Assuming the resonator supports only a single mode of the electromagnetic
field, its Hamiltonian is given by
\begin{equation}
H_r=\hbar\omega_r\left(a^\dag a+\frac{1}{2}\right),
\end{equation}
where $a^\dag$($a$) is the creation(annihilation) operator which
creates (destroys) a single photon in the cavity. The eigenstates
of the resonator described by this Hamiltonian are Fock states
$\vert 0\rangle \ldots \vert n\rangle \ldots$ with $n$ photons.
The single mode condition corresponds to a harmonic oscillator
with the energy levels at $\hbar\omega_r (n+1/2)$ and the
zero-point energy $\hbar\omega_r /2$.

The interaction between the radiation and the qubit is described as
the dipole interaction $H_g=-\mathbf{\hat{\mu}\cdot \hat{B}}$
between a magnetic moment $\mathbf{\mu}$ of the persistent current
circulating in the qubit loop and the magnetic field $\mathbf{B}$ in
the resonator.  Introducing quantization, this term in the Hamiltonian can be
written in the form
\begin{equation}
H_g=\hbar g(a^\dag \sigma^- + \sigma^+ a),
\end{equation}
where $\sigma^+$($\sigma^-$) is the raising(lowering) operator for
the qubit.  This expression is valid in the so-called
\textit{non-dispersive} regime where $\omega_r \approx \omega_0$
\footnote{In the dispersive regime the effective Hamiltonian becomes
$H=g^2/\Delta(\hat{\sigma}^+ \hat{\sigma}^- + \hat{a}^\dag
\hat{a}\hat{\sigma_z})$ \cite{gerry}.}. The constant $g$
characterizes the qubit-photon interaction strength and the
expression in the brackets describes the process whereby the qubit
can be excited by absorbing a photon, or a photon can be generated
at the expense of de-exciting the qubit into its ground state. In
the next section we shall explicitly calculate the dipole coupling
strength $g$ for specific designs of the qubit and the cavity.

The interaction of the coupled system with an external classical
drive field can be seen as a periodic exchange of photons between
the resonator and the driving field:
\begin{equation}
H_I=\xi (e^{-i\omega t}a^\dag+e^{i\omega t}a),
\end{equation}
where $\xi$ is the drive amplitude. One can also drive the qubit
directly using a separate control line  leading to terms $\xi '
(e^{-i\omega t}\sigma^+ +e^{i\omega t}\sigma^-)$, but we shall not
consider this case any further.

Adding the terms (2)-(5) together we arrive at the driven
Jaynes-Cummings Hamiltonian \cite{wallsandmilburn}
\begin{eqnarray}
H=\frac{\hbar\omega_0}{2}\sigma_z + \hbar\omega_r\left(a^\dag
a + \frac{1}{2}\right) + \hbar g(a^\dag \sigma^- + \sigma^+ a) \\
\nonumber +\: \xi (e^{-i\omega t}a^\dag + e^{i\omega t}a).
\label{eq:fullhamil}
\end{eqnarray}

\begin{figure}[b]
\centering
\includegraphics[width=6cm]{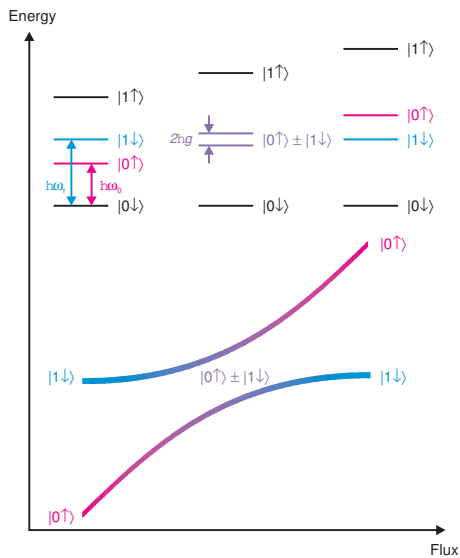}
\caption{\label{fig:EnergyLevels} Energy levels of the coupled
qubit-cavity system.  On the left of the diagram the qubit is far
detuned from the cavity.  As we move from left to right, the
magnetic flux threading the qubit loop is increased, tuning the
qubit transition frequency into resonance with the cavity.  On the
right of the diagram, the qubit is once again far detuned.}
\end{figure}

This Hamiltonian can be used to write down a master equation which
completely describes the dynamics of the system. All the
parameters which define the system can be conveniently written in
units of angular frequency -- a convention which we will follow
here. If the cavity is weakly coupled to an already weak drive
field one can achieve a regime when only two lower Fock states of
the resonator are relevant. Within the picture described above we
make one two-level system, the qubit, interact with another
two-level system, the resonator. When the qubit is detuned from
the resonator the eigenstates of the coupled system can be
written: $\vert0\downarrow\rangle$, $\vert0\uparrow\rangle$,
$\vert1\downarrow\rangle$ and $\vert1\uparrow\rangle$, where the
number represents a Fock state of the resonator and the arrow
represents the qubit state. However, when the qubit is brought
into resonance, $H_g$ couples the states $\vert0\uparrow\rangle$
and $\vert1\downarrow\rangle$ and lifts their degeneracy. The
system will oscillate between the states $\vert0\uparrow\rangle$
and $\vert1\downarrow\rangle$ at a frequency $\Omega = 2g$ , known
as the vacuum Rabi frequency \cite{semba},  giving rise to a
splitting \cite{alsing} of the central peak as shown in
Fig.~\ref{fig:EnergyLevels}. One can visualize this as a cycle in
which the resonator and qubit continuously exchange an amount of
energy equal to one photon. As the drive amplitude increases it
will start to perturb the system. This leads to a set of states
that are shifted by
\begin{equation}
E_n=\pm \sqrt{n} \hbar g [1-(2\xi/g)^2]^{3/4}.
\end{equation}
Hence, the effect of the drive is to \textit{reduce} the Rabi frequency.

The coupling strength $g$ can be determined experimentally by making
a spectroscopic measurement of the splitting $\Delta E_1$.  The
drive amplitude should be reduced until the splitting reaches its
maximum value, where it is equal to $2g\hbar$.

The above discussion covers the resonant regime, where the qubit is
tuned into resonance with the cavity.  In contrast, when the
detuning $\Delta=\omega_0-\omega_r$ is large, such that
$g/\Delta\ll1$, a dispersive Stark shift pulls the cavity frequency
by $\pm g^2/\Delta$. This so-called \textit{dispersive regime} can
be used to perform quantum non-demolition measurements of the qubit
state \cite{blais}.

The effects of the environment on the system are taken into account
by the term $H_E$. There are three types of damping that need to be
considered:
\begin{itemize}
    \item Photons leak out of the cavity at a rate $\kappa=\omega_r/Q$, where Q is the cavity quality factor.
    \item The qubit relaxes at a rate $\gamma=1/T_1$, where $T_1$ is the energy relaxation time.
    \item \textit{Pure} dephasing of the qubit at a rate
    $\gamma_\phi=1/T_\phi=1/T_2-2/T_1$, where $T_2$ is the
    dephasing time.
\end{itemize}
Dephasing plays a larger role in solid state systems than atomic
systems, due to the stronger interaction of solid state qubits with
their environments. In the absence of pure dephasing we would have $T_2=2T_1$ but in real
systems $T_2$ is frequently much shorter than that, indicating the need to take pure dephasing into account.

\section{Strong Coupling with a Flux Qubit}

Below we estimate the coupling strength $g$ using a semi-classical
approach.  We treat the flux qubit as a magnetic dipole and assume
that it is placed at a magnetic field antinode of the resonator.
The coupling strength is given by $g = \mu B_{0rms}/\hbar$, where
$B_{0rms}$ is the zero-point root mean square magnetic field
generated by the current fluctuations at the antinode of the
resonator.  The magnetic dipole moment of the qubit is given by
$\mu= I_p A$, where $I_p$ is the persistent current flowing around
the loop and $A$ is the loop area.  We can estimate $B_{0rms}$ by
considering the zero point energy of the resonator,
$\hbar\omega_r/2$. This energy cycles continuously between
inductive and capacitive components. The magnetic field is
determined by the inductive component, $LI(t)^2/2$, where $L$ is
the total equivalent inductance of the resonator near resonance
and $I(t)$ is the instantaneous current. At the moment when the
energy is purely inductive, we have $(1/2) LI_{max}^2 =
(1/2)\hbar\omega_r$. Since I(t) undergoes sinusoidal oscillation
we therefore have that
\begin{equation}
I_{rms} = \sqrt{\frac{\hbar\omega_r}{2L}}.
\end{equation}
Assuming that current flows in thin strips (whose width is
determined by the superconducting penetration depth) at the edges of
the centre conductor and ground plane, the
field at the antinode of the fundamental mode is approximately given by
\begin{equation}
B_{0rms} \approx \frac{\mu_0 I_{rms}}{\pi r},
\end{equation}
where $\mu_0$ is the permeability of free space and $r$ is half the
width of the gap between the centre conductor and the ground plane
(we assume the qubit is placed at the centre of the gap). Therefore,
the coupling strength between the qubit and resonator is given approximately by
\begin{equation}
g \approx \frac{I_p A \mu_0}{\hbar\pi r}
\sqrt\frac{\hbar\omega_r}{2L}.
\end{equation}
By inserting realistic values for the parameters in the above
equation we can obtain an estimate of $g$.

First, we choose the fundamental frequency of the resonator.  It
is convenient to choose a value that lies within the range 4--8
GHz, as this is well within the design scope of both the qubit and
resonator, and can be accessed with commercial microwave sources
and components. We choose $\omega_r/2\pi = 6$ GHz.  With a centre
conductor of width $\sim 10$~$\mu$m and a gap of width $\sim 5$~
$\mu$m, it is possible to achieve a total inductance $L \sim
2$~nH for a resonator operated at its fundamental frequency..

Next, we choose $I_p$ such that the transition frequency $\omega_0$
of the qubit at the degeneracy point is slightly less than that of
the resonator.  This will enable us to tune the qubit in and out of
resonance with the resonator by changing the external flux $\Phi_x$
threading the qubit loop. For the 3-junction persistent current qubit having two
junctions of critical current $I_c = 800$~nA and junction
capacitance $C = 4$ fF, and one junction of critical current $\alpha
I_c$, where $\alpha = 0.72$, we get $\omega_0/2\pi = 4.9$~GHz at
the degeneracy point and $I_p \approx 580$~nA. These parameters were obtained by
solving the Schroedinger equation numerically.
The transition frequency $\omega_0$ does not depend on the area $A$ of the qubit
loop, provided that the loop inductance remains small compared with
the Josephson inductance. However, we note that the larger the loop
area, the greater the (undesired) coupling to the environment.  Here
we choose a value $A \approx 8 $ $\mu$m$^2$.

With the above parameters we obtain $g/2\pi \approx 35$~MHz.  We
now compare this with the rate of photon loss from the resonator
$\kappa$ and the relaxation rate of the qubit $\gamma$.  When $g >
\kappa, \gamma$, the coupled system is able to undergo many cycles
($\approx 2g/(\kappa+\gamma)$) of vacuum Rabi oscillation before
losing coherence.  This is important for applications such as
single microwave photon generation. The photon loss rate from the
resonator is given by $\kappa/2\pi = \omega_r/Q$, where $Q$ is the
loaded quality factor of the resonator.  It is possible to design
a resonator with $Q = 10^5$, yielding $\kappa/2\pi \approx 0.1$
MHz.  The relaxation rate of the qubit is given by $\gamma =
2\pi/T_1$. Taking $T_1 \sim 1$ $\mu$s \cite{chiorescu}, we obtain
$\gamma/2\pi \approx 1$ MHz.
Naturally, the values for $T_1$
and $T_2$ for a real system can not be predicted with any accuracy and will depend on the experimental
conditions. However, based on the data available in the literature \cite{bertet,kakuyanagi}
we believe that the aforementioned values are reasonable. Hence,
it is clear that our estimated value of $g$ for the
persistent current qubit should satisfy the strong coupling criterion.

For an RF SQUID with critical current $I_c = 10$ $\mu$A, area $A =
64$ $\mu$m$^2$, loop inductance $L_{SQUID} = 35$ pH and junction
capacitance $C = 50$ fF, we obtain a transition frequency
$\omega_0/2\pi = 4.6$ GHz. In contrast to the persistent current
qubit, the area of the RF SQUID does affect the transition
frequency, via the loop inductance. It is difficult to reduce the
area further than the value we have chosen, as this necessitates
increasing the critical current and decreasing the junction
capacitance, which becomes increasingly difficult to achieve in
practice.  Close to the degeneracy point, the persistent current
$I_p$ in the above SQUID is expected to be about 5 $\mu$A.
Combined with the increased loop area, this is likely to lead to
an even larger coupling strength $g$ than predicted for the
persistent current qubit (unless the SQUID is displaced
significantly from the antinode of the resonator) but at the same
time make qubit more susceptible to noise. The fact that the loop
area of the 3-junction persistent current qubit can be made small
enough to render it relatively insensitive to flux noise is one
reason why it has been so successfully used by e.g Mooij and
co-workers \cite{Mooij-1999,chiorescu}.

\section{Numerical simulation of the spectrum under microwave excitation}

Having shown that it is possible to reach the strong coupling
regime with a flux qubit coupled to a coplanar resonator, we now
simulate the results of a spectroscopic experiment to measure $g$.
The experiment would involve driving the coupled system with an
external microwave field whose frequency $\omega_l$ would be swept
through the resonance of the coupled system.  The most
straightforward way to probe the response of the system would be
to use what effectively amounts to a standard microwave
transmission ( S$_{12}$ ) measurement of the cavity. The
experiment would be done by starting with the qubit far detuned
from the resonator, then stepping the external magnetic flux to
tune the qubit through the cavity resonance. This type of
experiment would allow us to record the output power of the
resonator as a function of qubit Larmor frequency.

These simulations were performed by solving the master equation
using a Liouvillian with the Hamiltonian (\ref{eq:fullhamil}) and
three collapse (Lindblad) operators which account for the decay and
dephasing of the qubit and the cavity at the aforementioned rates
$\kappa$,$\gamma$ and $\gamma_\phi$ (i.e. the effects of $H_E$). A
brief description of the formalism can be found in appendix
\ref{appendix:dissipation}. All simulations were performed  using
the ``Quantum Optics Toolbox" developed by Tan\cite{Tan}. Note that all the
figures show the \textit{spectrum} of the intracavity field (see appendix
\ref{ap:spectrum} for the definition and a description of how it is
calculated).

In the regime where the qubit Larmor frequency $\omega_0$ is very
far detuned from the cavity (i.e. when even dispersive effects are
negligible) the qubit and the cavity are effectively decoupled and
no exchange of energy can take place. If the system is probed by
measuring the response of the cavity,  a single peak located
at the bare resonance frequency $\omega_r$ will be seen.

\begin{figure}
\centering
\includegraphics[width=7.5cm]{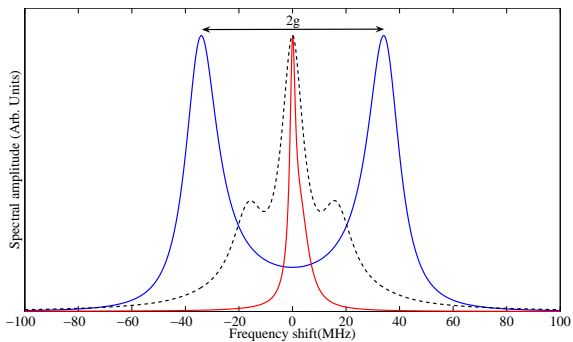}
\caption{\label{fig:1D_spectrum}  The spectrum far from resonance (where the only effect
of the qubit is to broaden the resonance) and at zero de-tuning for small drive (mean steady-state photon
number $\langle n \rangle < 10^{-3}$), the latter giving rise to a
Rabi splitting $2g$. Shown is also the spectrum at large ($\langle
n \rangle \approx 8$) drive amplitudes (\textit{dashed line}).}.
\end{figure}

\begin{figure}
\includegraphics[width=\columnwidth]{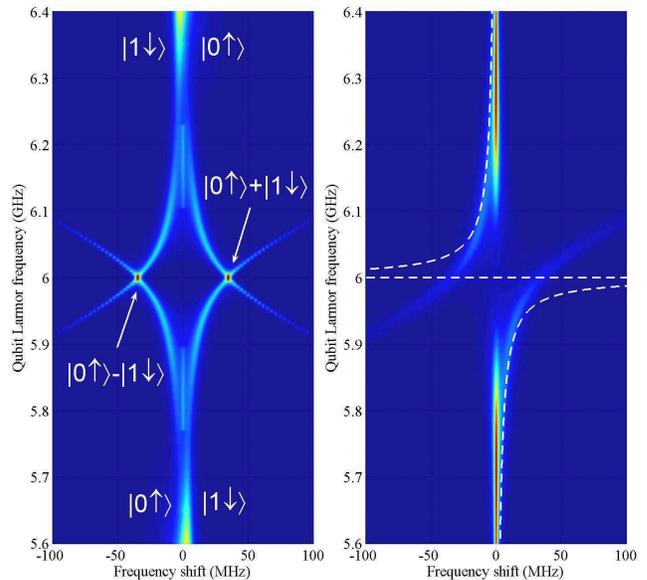}
\caption{\label{fig:power_spectrum}  (Color online) Power spectrum
of the coupled qubit-resonator system as a function of qubit
detuning in the strong coupling limit. The frequency of the drive
field $\omega_l$ is held at the resonance frequency $\omega_r$ of
the bare resonator, while the Larmor frequency $\omega_0$ of the
qubit is tuned by changing the external magnetic field.
\textit{Left}: Spectrum in the absence of pure dephasing.
\textit{Right:} Adding a pure dephasing channel to the dissipation
results in an asymmetric spectrum, here the pure dephasing rate
$\gamma_\phi$ is 9.5~MHz. Shown is also one branch of the
expression $\pm g^2/\Delta$ \textit{(white dashed line)}. The
following parameters were used in the simulations (in GHz):
$\omega_r=\omega_l=6$, $g=0.035$, $\kappa=0.004$, $\gamma=0.001$
and $\xi/\hbar=0.25\kappa$. }
\end{figure}

If instead the qubit is tuned exactly on resonance ($\Delta=0$)
the effects of the coupling become clearly visible. Now, there are
two peaks located at $\omega_r\pm \Omega/2=\omega_r \pm g$ as can
be seen in fig. \ref{fig:1D_spectrum}.
In between these two extremes there is a gradual change from a
single- to a double-peaked spectrum where the splitting is
approximately $g^2/\Delta$ as can be seen in the left picture of
fig. \ref{fig:power_spectrum}. In this case the pure dephasing rate $\gamma_\phi$ was set to zero.
The result is a "diamond-shaped" picture with a maximum splitting
of $2g$ at zero detuning.

Far from resonance  we can identify the four branches as being
associated with the states $|0 \uparrow\rangle$ and $|1
\downarrow\rangle$ above and below the resonance. Exactly on
resonance the system is in a superposition of states $|0
\uparrow\rangle \pm |1 \downarrow\rangle$. This is in agreement
with the diagram shown in fig.~\ref{fig:EnergyLevels}. Note that
e.g. an interferometric measurement method \cite{Oxborrow-2005}
must be used in order to be able to observe the whole spectrum, a
simpler transmission experiments would only see one peak
off-resonance since such a measurement only records transitions
between photon states where the final state can emit a photon; in
this case $|0\uparrow\rangle \leftrightarrow |1\downarrow\rangle
$.

The right picture in fig.~\ref{fig:power_spectrum} shows the
spectrum when pure dephasing is introduced to the model. The result
is somewhat more complicated in that the spectral weights are
asymmetric with respect to $w_r-w_l=0$ when the qubit is detuned
from resonance. The reason for this is that dephasing leads to a loss of
coherence, meaning that the qubit tends to stay in its ground state and the coupling between the states  $|0\uparrow\rangle$
and $|1 \downarrow\rangle$ is reduced, the system is therefore
no longer in a superposition of those states. Whence, states with zero photons
in the cavity are effectively "decoupled" from the one-photon states. Since only states which allow
for (at least) one photon in the cavity can be measured it follows that only the two branches (one above and one below resonance) that
(approximately) correspond to $\pm |1\downarrow\rangle$
will be clearly visible. Note, however, that despite the loss of
coherence an on-resonance measurement would still show two peaks
separated by $2g$, i.e. exactly on resonance this situation is
effectively indistinguishable from the more coherent case even when the full spectrum is measured.

While there are no bound states in the limit of strong driving
$\xi>g/2$ a continuum of states still exists giving rise to
complex spectra (dashed line in fig.~\ref{fig:1D_spectrum}). The
structure is reminiscent to the so-called "Mollow" peaks, well
known in atomic physics from e.g. fluorescence spectroscopy.
However, in the latter case the peaks are the result of strong
driving of the \textit{atom} (qubit) whereas in this simulation
the \textit{cavity} is being driven so the similarity is somewhat
superficial. When as in this case the cavity is driven so strongly
that there are on average several photons in the cavity, we see
both a drive induced shift of the position of the sidebands and a
reappearance of the central peak. Note that since it is difficult
to directly relate the parameter $\xi$ to the power output from a
microwave generator, care must be taken not to drive the system
inadvertently into this regime.

For the persistent current qubit considered here, a detuning of
$\pm 0.4$ GHz corresponds to an external magnetic flux in the
range $\pm 10^{-4} \Phi_0$ which is a useful value for a real
experiment. However, for the RF SQUID, the flux required is rather
less: $\pm3\cdot 10^{-5} \Phi_0$.

One effect which is not taken into account in our simulations is
the presence of thermal photons in the cavity. Thermal effects can,
in general, be neglected when working with optical cavities due to
the very small average number of photons at those frequencies. In
experiment on solid state qubits this is, however, not generally
true since they are operated in the microwave range. Also, the
relevant temperature scale is set not only by the the phonon
temperature (e.g. the temperature of the mixing chamber of a
dilution refrigerator) but also by the amount of noise
(essentially "hot" photons) which reaches the system via the
leads. However, in a well-filtered system a total temperature of
50 mK is attainable. The resonator will then nearly be in its
ground state with an average thermal occupancy $\bar{n}$ of 0.009
and thermal fluctuations in the photon number of the order of 0.1.
This justifies ignoring thermal effects in our simulations for
now. That said, even a moderate increase in temperature can
significantly change the outcome of an experiment \cite{rau}. One
further simplifying assumption in the model is that $T_1$ and
$T_2$ do not change as the qubit is detuned from the optimal bias
point $\Phi_0/2$. While this is clearly unrealistic, it has been
experimentally shown \cite{kakuyanagi} that neither parameter
should change dramatically in the parameter range considered here,
giving some justification to this approximation.

\section{Applications of circuit-QED}

One of the most important applications of circuit QED is the
generation of single microwave photons on demand.  Single photon
sources in the optical regime have been realized using e.g. cavity
QED with atoms and high finesse optical cavities
\cite{Oxborrow-2005}. Design and fabrication of deterministic
sources that operate in the microwave regime have proved to be
more difficult, but a source based on superconducting circuit-QED
was recently demonstrated \cite{houck}. This kind of source could
be used for quantum radiometry, as well as for quantum information
applications such as quantum key distribution.

Various schemes can be envisaged
\cite{mariantoni,marquardt,zagoskin,saito}, by which single
photons can be generated with a circuit QED device. Below, we
describe a straightforward technique, based on manipulation of the
qubit state with microwave pulses and rapid changes in the DC
magnetic field to tune the qubit in and out of resonance with the
cavity.

\begin{figure}
\begin{center}
\includegraphics[width=70mm]{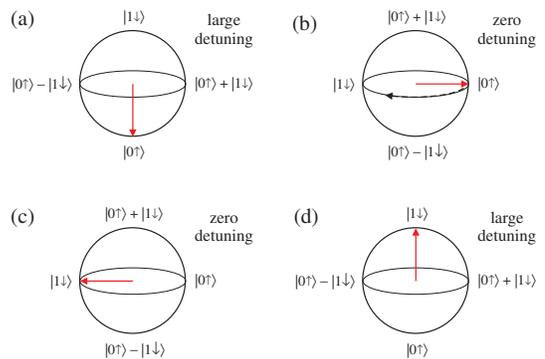}
\caption{(Color online). Bloch sphere diagrams showing the state of
the qubit-cavity system at successive stages of the single photon
generation process.}
\label{fig:SPG}
\end{center}
\end{figure}

Our technique begins with the qubit far detuned from the cavity
and the combined system in its ground state $\vert
0\downarrow\rangle$. A microwave $\pi$-pulse is applied to the
qubit to excite it to the state $\vert 0\uparrow\rangle$ [Fig.\
\ref{fig:SPG}(a)], and this is followed by a step in the magnetic
field to bring the qubit into resonance with the cavity [Fig.\
\ref{fig:SPG}(b)]. The state of the system immediately after the
step is still $\vert 0\uparrow\rangle$, but due to the
qubit-cavity interaction on resonance, this is no longer an
eigenstate, so the system begins to precess around the equator of
the Bloch sphere at the vacuum Rabi frequency $2g$. After a time
$2\pi/4g$, the state of the system will be $\vert
1\downarrow\rangle$ [Fig.\ \ref{fig:SPG}(c)].  This means that
coherent energy exchange has taken place between the qubit and the
cavity, creating a photon-like state.  Another step in magnetic
field detunes the qubit from the cavity so that the state $\vert
1\downarrow\rangle$ is once more an eigenstate of the system
[Fig.\ \ref{fig:SPG}(d)]. The system remains in this state until
the photon decays out of the cavity into one of the external
waveguides in a time of order $2\pi/\kappa$.  By repeating this
sequence many times, photons can be generated on demand, provided
that the time window within which they are required is much longer
than $2\pi/\kappa$.

If a scheme similar to the one above is implemented, it is important
to prove that it generates single photons deterministically, rather
than stochastically. At optical frequencies this is done by studying
photon-counting statistics using interferometric measurements.  Such
measurements require the use of a beamsplitter. An analogous
experiment can be envisaged in the microwave regime, provided that a
microwave beamsplitter can be realised. Such a device has been
proposed recently \cite{mariantoni}, and could lead to a microwave
analogue of the Hanbury-Brown and Twiss interferometer
\cite{Oxborrow-2005}.

\section{Conclusions}

We have shown that it should be possible to reach the strong
coupling regime using a flux qubit coupled to a coplanar waveguide
resonator. If realized, it would open the door to potential
applications in metrology, quantum communication and experimental
tests of quantum mechanics.  The fact that conventional
lithography can be used to fabricate the samples and that the
experimental parameters can be chosen freely can in some cases a
be significant advantage compared to CQED implentantaions
utilizing e.g. atoms of ions. Since the relevant frequencies are
in the microwave regime it is also possible to use well
established methods to manipulate the system. The main drawback
compared to experiments done at optical frequencies is the short
coherence time of the qubit and the fact that the system must be
operated at very low temperatures.

\section{Acknowledgments}
The authors would like to thank Mark Oxborrow, Alexandre Zagoskin,
Alexander Blais and Vladimir Antonov for helpful discussions and
comments. We would also like to thank Scott Parkins for his help
with the Quantum Optics Toolbox. This work was funded by the UK
Department of Trade and Industry Quantum Metrology Programme,
project QM04.3.4. and the Swedish Research Council.

\appendix

\section{Derivation of the Hamiltonian}
It is useful to compare the informal procedure used in the
introduction of this paper to derive the Hamiltonian with a more
formal approach. Starting with the bare qubit Hamiltonian
$-\frac{1}{2} (\epsilon \sigma_z + \Delta \sigma_x)$ where
$\epsilon=2I_p(\Phi_x-\Phi_0/2)$ we proceed just as before by
noting that the flux threading the qubit loop will be modulated
via the mutual inductance $M$ that couple the fluctuations in the
cavity to the qubit. Writing the total external flux as
$\Phi_x=\Phi_x^{DC}+\delta \Phi$, where $\delta
\Phi=M\sqrt{\frac{\hbar \omega_r}{2L}} (a^\dag+a)$, adding the
Hamiltonian for the oscillator mode and the external field and
finally transforming into the eigenbasis of the qubit we get
\begin{eqnarray}
H=&& \hbar \omega_r \left( a^\dag a+\frac{1}{2} \right)+
\frac{\hbar \omega_0}{2}\sigma_z  +\hbar g (a^\dag\sigma_-
+\sigma_+a) \sin \theta \nonumber \\ &&+\xi (e^{-i\omega
t}a^\dag+e^{i\omega t}a)-\hbar g (a^\dag+a)\sigma_z \cos \theta
\label{eq:comhamil}
\end{eqnarray}
in the RWA. Here we have introduced the mixing angle
$\theta=\arctan{\Delta/\epsilon}$. This Hamiltonian is identical
to the J-C Hamiltonian (\ref{eq:fullhamil}) except that we now
have an effective coupling $g \sin \theta$ and an extra term
$\hbar g (a^\dag+a)\sigma_z \cos \theta$ which is zero when the
qubit is operated at the degeneracy point $\theta=\pi/2$. By
moving to an interaction frame rotating at the drive frequency
$\omega$ we see that all terms in the Hamiltonian
(\ref{eq:comhamil}) are time-independent except the last term
which picks up a factor $\exp(-iw\omega)$, meaning it can be
neglected in the rotating wave approxiamtion. Note, however, that
this additional term can potentially play a role in the dispersive
regime.

\section{Dissipation}
\label{appendix:dissipation}
The effects of the environment on a quantum system is in general
very difficult to model but is nevertheless crucial to understand
since it is the cause of decoherence. However, assuming the interaction
with the environment is \textit{Markovian} the evolution of the (reduced)
density matrix of the system can be described by a master equation
$\dot{\rho}=\mathbf{L} \rho$ of Lindblad form \cite{breuer}
\begin{equation}
\frac{\partial \rho}{\partial t}= -\frac{i}{\hbar}[H,\rho]
+\sum^{3}_{k=1} \left(C_k \rho C_k^\dag -\frac{1}{2}
\left(C_k^\dag C_k \rho+ \rho C_k^\dag C_k \right)\right)
\label{eq:lindbladform}
\end{equation}
where $C_k$ are Lindblad operators. In the case considered here we
have 3 Lindblad operators. Firstly, the relaxation from the
excited state to the ground state at a rate $\gamma_1=1/T_1$
represented by a Lindblad operator proportional to the lowering
operator $\hat \sigma^-$, i.e. $C_1=\sqrt{\gamma_1} \hat
\sigma^-$. The cavity is loosing energy at a rate
$\kappa=\omega_r/Q$ which leads to the "destruction" of photons in
the system,  $C_2=\sqrt{\kappa} a$. Finally, we also need to
consider \textit{pure} dephasing of the qubit  at a rate
$\gamma_\phi=1/T_\phi=1/T_2-1/2T_1$ where $T_2$ is the usual total
dephasing time of the qubit. This process is represented by the
operator $C_3=\sqrt{\frac{\gamma_\phi}{2}}  \hat\sigma_z$.

\section{Calculation of the spectrum}
\label{ap:spectrum} Our aim is to calculate the steady-state power
spectrum $S(\omega)$ of the intracavity field, formally this is
defined in terms of the photocount output from the cavity as seen
by a monochromatic detector \cite{wallsandmilburn}. The spectrum
can be calculated from the 2-time correlation function
\cite{glauber} $\langle a^\dag(t+\tau) a(\tau)\rangle$
\begin{equation}
S(\omega)=\frac{1}{2\pi} \int_{-\infty}^{\infty} e^{-i \omega
\tau} \langle a^\dag(\tau+t) a(t) \rangle d \tau
\end{equation}
which can be evaluated using the quantum-regression theorem
\begin{equation} <a^\dag(\tau+t) a(t)>=\mathrm{Tr}\{a^\dag
e^{\mathbf{L}\tau} a \rho\}
\end{equation}
 where the Liovillian
(which includes the three Lindblad operators defined in Appendix
\ref{appendix:dissipation}) is given by the right hand side of
equation \ref{eq:lindbladform} and $\rho$ is the steady-state
density matrix which is the solution to $\mathbf{L} \rho=0$ These
calculations are straightforward using the built-in routines of
the "Quantum Optics Toolbox".
\bibliographystyle{unsrt}
\newpage
\bibliography{CQED_metrology}

\end{document}